\documentclass[12pt, a4paper]{article}
\usepackage{a4wide}
\usepackage{epsf}
\begin{document}
\newcommand{\qed}{\hfill \rule{2mm}{2mm}}
\newcommand{\pf}{{\bf Proof : }}
\newtheorem{definition}{Definition}
\newtheorem{construction}{Construction}
\newtheorem{theorem}{Theorem}
\newtheorem{question}{Question}
\newtheorem{lemma}{Lemma}
\newtheorem{proposition}{Proposition}
\newtheorem{remark}{Remark}
\newtheorem{corollary}{Corollary}
\newtheorem{example}{Example}
\newcommand{\binom}[2] {\mbox{$\left( { #1 \atop #2 } \right)$}}

\title{On the non-existence of a universal Hadamard gate} 
\author{
Preeti Parashar\\
Physics and Applied Mathematics Unit\\
Indian Statistical Institute\\
203 B T Road, Kolkata 700 108\\
Email: parashar@isical.ac.in
}
\date{}
\maketitle

\begin{abstract}
We establish the non-existence of a universal
Hadamard gate for arbitrary unknown qubits, by considering two different 
principles; namely,
no-superluminal signalling and non-increase of entanglement 
under LOCC. 
It is also shown that these principles are not violated if and only if
the qubit states are of the special form 
obtained in our recent work [IJQI (in press), quant-ph/0505068].
\end{abstract}



The Hadamard gate is a one qubit unitary operator which has played
a very crucial role in the developement of quantum algorithms 
\cite{d, dj, sh}. 
It rotates the computational basis (CB) 
states $|0\rangle$ and $|1\rangle$ by
creating an equal superposition of the amplitudes, 
and the new states so obtained are orthogonal to each other.
Can this action of the Hadamard gate be generalized to any 
arbitrary unknown qubit state? 
It was proved in \cite{pati1} using unitarity of the evolution, that
there does not exist a Hadamard gate for a
completely arbitrary qubit state. In other words, the Hadamard gate
is not universal.
Then the natural question to ask is: What is the most general 
class of states for which it is  
possible to design this gate in a universal manner? 
We endeavored to address this aspect in our recent investigation
\cite{mp} where a special ensemble of qubits
was obtained. The Hadamard gate
holds for each and every state belonging to this ensemble.

The restrictions imposed by the structure of quantum mechanics 
has led to several other {\em impossible}
operations in quantum information theory; for example, the famous
no-cloning theorem \cite{wz} and the no-deleting principle \cite{pb1}.
There have also been attempts to derive them by applying some other 
fundamental principles of nature, like no-faster-than-light communication 
between spatially separated parties \cite{gisin, hs, pati2, pb2} and 
preservation of entanglement for closed systems under local operations 
\cite{horo}.
Another important no-go theorem is the non-existence of a universal 
flipping operator (NOT gate) \cite{bh} which would take an 
arbitrary qubit state to its orthogonal complement. An alternative proof 
of this was recently given in \cite{kar} from the
constraint of no-signalling and also from the consideration that 
the amount of entanglement shared between two spatialy separated parties 
cannot be increased by local operations and classical communication 
(LOCC). These pursuits serve as the chief motivation to undertake the 
present investigation.
                            
The purpose of this communication is two-fold. Firstly, we  
prove the non-existence of a universal Hadamard
gate  by imposing (separately) the two fundamental physical 
principles mentioned above; namely, $(i)$ the 
no-superluminal signalling condition of special relativity, and $(ii)$
non-increase of entanglement under LOCC. Secondly, 
we show that as a consequence of non-violation of these two laws,
the qubit states {\em must} be of a  
particular type. This gives the largest set of states for  
which the Hadamard gate is valid, and matches exactly with the result 
derived in \cite{mp} from an altogether different consideration. 

Let us take the CB states $|0\rangle$, $|1\rangle$ and a
third state $|\psi\rangle$ which is an 
arbitrary linear superposition of these two. Thus,
$|\psi\rangle = a |0\rangle + b |1\rangle$, 
where $a$ and $b$ are {\em non-zero} complex numbers satisfying the 
normalization condition $a^*a + b^*b = 1$.
We define the action of the unitary Hadamard operator $H$ on these 
states as follows:
\begin{equation}
\label{h1}
H|0\rangle = \frac{1}{\sqrt{2}}(|0\rangle + |1\rangle), ~~
H|1\rangle = \frac{1}{\sqrt{2}}(|0\rangle - |1\rangle), ~~    
H|\psi\rangle = \frac{1}{\sqrt{2}}(|\psi\rangle + |{\overline 
\psi}\rangle),
\end{equation}
where 
$|{\overline \psi}\rangle = b^* |0\rangle - a^* |1\rangle$ is the 
orthogonal complement of $|\psi\rangle$. The Hadamard gate is explicitly
represented by the matrix 
$H = \frac{1}{\sqrt{2}}
\left[\begin{array}{cr}
1 & 1 \\
1 & -1 \\
\end{array}\right]$.
The transformation of
$|{\overline \psi}\rangle$
can be easily deduced from that of 
$|\psi\rangle$ in (\ref{h1}) and reads as 
$H|{\overline \psi}\rangle = \frac{1}{\sqrt{2}}(|\psi\rangle -
|{\overline \psi}\rangle)$.         

{\bf No-superluminal signalling:}
First we shall show that the Hadamard operation defined in 
(\ref{h1}) implies signalling. For this purpose                                                                           
let us consider that Alice and Bob share the entangled state 
\begin{equation}
\label{ent1}
\phi_{AB} = \frac{1}{2} \left(|0\rangle_A |0\rangle_B + |1\rangle_A 
|\psi\rangle_B + 
|2\rangle_A |1\rangle_B + |3\rangle_A |{\overline \psi}\rangle_B \right),
\end{equation}
where Alice's particle is four-dimensional and Bob's is 
two-dimensional. 

The density matrix of the combined system is $\rho_{AB} = 
|\phi_{AB}\rangle 
\langle \phi_{AB}|$. Alice's reduced density matrix can be obtained 
by tracing out Bob's part. So
\begin{eqnarray}
\label{ra1}
\rho_{A} = tr_B(\rho_{AB})& = & \frac{1}{4}[ |0\rangle \langle 0| + 
a |1\rangle \langle 0| + b^* |3\rangle \langle 0| + a^* |0\rangle \langle 1|
 + |1\rangle \langle 1| + b^* |2\rangle \langle 1| \nonumber \\
& + & b |1\rangle \langle 
2| + |2\rangle \langle 2| - a^* |3\rangle \langle 2| + 
b |0\rangle \langle 3| - a |2\rangle \langle 3| + |3\rangle \langle 3|].
\end{eqnarray}
Now Bob applies Hadamard transformation on his qubit in Eq(\ref{ent1})
but does not communicate anything to Alice. The shared 
state thus changes to 
\begin{eqnarray}
(I \otimes H) \phi_{AB} = \phi'_{AB} 
= \frac{1}{2\sqrt{2}} &[& |0\rangle |0\rangle + |0\rangle |1\rangle + 
|1\rangle |\psi\rangle + |1\rangle |{\overline \psi}\rangle\nonumber \\ 
&+& |2\rangle |0\rangle - |2\rangle |1\rangle + |3\rangle |\psi\rangle - 
|3\rangle |{\overline \psi}\rangle].
\end{eqnarray}
After this operation, Alice's new reduced density matrix becomes 
\begin{eqnarray}
\label{ra2}
\rho'_A = tr_B(\rho'_{AB})
&=& \frac{1}{8} [ 2( |0\rangle \langle 0|+ |1\rangle \langle 1|+ |2\rangle 
\langle 2|+ |3\rangle \langle 3| ) \nonumber \\
&+& (a + b^* + b - a^*) |1\rangle \langle 0| 
+(a - b^* + b + a^*) |3\rangle \langle 0| \nonumber \\
&+& (a^* + b^* + b - a) |0\rangle \langle 1| +
(a^* - b^* + b + a) |2\rangle \langle 1| \nonumber \\
&+& (a + b^* - b + a^*) |1\rangle \langle 2| 
+ (a - b^* - b - a^*) |3\rangle \langle 2| \nonumber \\
&+& (a + b^* - b + a^*) |0\rangle \langle 3| +
(a^* - b^* - b - a) |2\rangle \langle 3|].
\end{eqnarray}
Comparing the coefficients of each term in (\ref{ra1}) and (\ref{ra2}),
it is clear that $\rho'_A \ne \rho_A$ for arbitrary choices of the 
parameters $a$ and $b$.
So, in principle, Alice can distinguish between $\rho_A$ and 
$\rho'_A$, although Bob has not informed her anything about his operation. 
This implies that superluminal communication  has taken place 
with the help of  quantum non-local resourse (entanglement).
But special theory of relativity forbids faster-than-light 
communication. Hence, we conclude that
Hadamard gate does not exist for an arbitrary qubit
as it leads to signalling. 

If however the no-signalling constraint is imposed, then 
$\rho_A$ and $\rho'_A$ should be equal because the action of $H$ is a 
trace preserving local operation performed only at Bob's side.
Since $a$ and $b$ are complex, we can write $a = \alpha_a + i \beta_a$, 
$b = \alpha_b + i \beta_b$, where $\alpha_a, \alpha_b, \beta_a$ and 
$\beta_b$ are all real.
So $\rho'_A = \rho_A$ implies $a + b^* + b - a^* = 2a$, which gives,
$\alpha_b = \alpha_a$. 
Further, $a - b^* + b + a^* = 2b^*$ yields
$2 \alpha_a = 2\alpha_b - 4i\beta_b$, i.e., $\beta_b = 0$, implying 
that $b$ is purely real.
Letting $\alpha_a = \alpha,~ \beta_a = \beta$, we find that 
the qubit state assumes the form
\begin{equation}
\label{rq}
|\psi\rangle = (\alpha + i \beta) |0\rangle + \alpha |1 \rangle,
\end{equation}
with the normalization condition
$2 \alpha^2 + \beta^2 = 1$. This peculiar form has one complex
and one pure real amplitude (in CB); the two real parts being equal.
Interestingly, this is exactly the same ensemble that 
was obtained in \cite{mp}.  
It is in fact the largest set of qubit states (along with $|0\rangle$) for 
which the Hadamard gate 
can be designed universally. On the other hand,
$|\overline{\psi}\rangle = \alpha |0\rangle - (\alpha - i \beta) 
|1\rangle$ (along with $|1\rangle$) is the set of orthogonal complements.
These two sets formed nice trajectories when represented on the bloch 
sphere.\\

{\bf Non-increase of entanglement under LOCC:}
Next, we show the non-validity of the Hadamard transformation 
(\ref{h1}) for a general qubit, by considering the fact that local
operations and classical communication cannot increase the entanglement 
content of a quantum system. Unfortunately,
the resource state shared earlier, cannot be used in this case. 
There is no change in entanglement before and after the Hadamard operation 
since the eigenvalues of
$\rho_A$ and $\rho'_A$ are equal.
Therefore, let us consider a different shared state which is of the 
form \cite{kar}
\begin{equation}
\label{sep1}
{|\Phi\rangle}_{AB} = \frac{1}{1 + b^*b} [|0\rangle_A 
\frac{{|0\rangle}_{B1}
{|1\rangle}_{B2} - {|1\rangle}_{B1} {|0\rangle}_{B2}}{\sqrt{2}}
+ {|1\rangle}_A \frac{{|0\rangle}_{B1}
{|\psi\rangle}_{B2} - {|\psi\rangle}_{B1} {|0\rangle}_{B2}}{\sqrt{2}}],    
\end{equation}
where the first qubit is with Alice while the other two are with Bob.
This state is a product state in the A:B cut since it can be written as 
\begin{equation}
\label{sep2}
{|\Phi\rangle}_{AB} = \frac{1}{1 + b^*b} [\{|0\rangle_A + b |1\rangle_A \}
\otimes \{ \frac{{|0\rangle}_{B1}
{|1\rangle}_{B2} - {|1\rangle}_{B1} {|0\rangle}_{B2}}{\sqrt{2}} \}].
\end{equation}                                                         
Following the earlier protocol, we find Alice's reduced density 
operator
\begin{equation}
\rho_A = \frac{1}{1+b^*b} [ |0\rangle \langle 0| +
b^* |0\rangle \langle 1| + b |1\rangle \langle 0|
+ b^*b |1\rangle \langle 1| ].
\end{equation}                                                     
The eigenvalues of this matrix are $0$ and $1$. Obviously, the
amount of entanglement given by the van Neumann entropy is zero. 
Now Bob applies the trace preserving Hadamard transformation defined
in (\ref{h1}) 
on the last particle $(B2)$ in Eq(\ref{sep1}), which results in the state 
\begin{eqnarray}
\label{enh}
|\Phi'\rangle_{AB} = \frac{1}{2\sqrt{N}} &[& |000\rangle - |001\rangle - 
|010\rangle + |011\rangle \nonumber \\
&+& |10\psi\rangle + |10{\overline \psi}\rangle - |1\psi0\rangle - 
|1\psi1\rangle ],
\end{eqnarray}
where $N = 2 + \frac{1}{4}((a-a^*)^2 - (a+a^*)(b+b^*))$.
Since $a$ and $b$ are arbitrary, so in general, the above state is 
entangled in the A:B cut. This implies 
that entanglement has been created by local operation. 
However, we know that entanglement cannot be increased by local operations
even if classical communication is allowed.
Therefore, we conclude that Hadamard transformation of an 
arbitrary, unknown qubit is an invalid operation. 

Our next task is to derive the conditions under which 
the entanglement in the state would remain the same (zero in this case) 
before and after the 
application of the Hadamard gate. For this purpose we have to compare 
the eigenvalues
of the respective density matrices on Alice's side. So after Bob's 
operation
\begin{eqnarray}
{\rho'}_A = \frac{1}{4N} [&4& |0\rangle \langle 0| + 
(a + a^* + b + b^*) |0\rangle \langle 1| + 
(a + a^* + b + b^*) |1\rangle \langle 0|\nonumber \\ 
&+& (4 + (a - a^*)^2 - (a + a^*)(b + b^*)) |1\rangle \langle 1|]. 
\end{eqnarray}
It is, however, more convenient to express $a$ and $b$ in terms of real 
and imaginary components as follows:                                            
\begin{equation}
{\rho'}_A = \frac{1}{N} [|0\rangle \langle 0| +
\frac{\alpha_a + \alpha_b}{2} |0\rangle \langle 1| +
\frac{\alpha_a + \alpha_b}{2} |1\rangle \langle 0| +
( \alpha_a^2 + \alpha_b^2 + \beta_b^2 - \alpha_a \alpha_b) |1\rangle 
\langle 1|],
\end{equation}        
where $N = 1+ \alpha_a^2 + \alpha_b^2 + \beta_b^2 -
\alpha_a \alpha_b$.
The eigenvalue equation of the above matrix gives two roots:
\begin{equation}
\label{ro}
\lambda_{\pm} = \frac{1}{2} \pm \frac{\sqrt{(N-2)^2  + 
(\alpha_a + \alpha_b)^2}}{2N} 
\end{equation}
In order to maintain the same amount of entanglement in the system, 
we equate these two roots of $\rho'_A$ to the eigenvalues of $\rho_A$. 
This yields the constraint:
\begin{equation}
\label{co}
\beta_b^2 = -\frac{3}{4} (\alpha_a - \alpha_b)^2.
\end{equation} 
But since $\alpha_a$ and $\alpha_b$ are both real, so the
squared term on r.h.s. will always be positive.
This implies that $\beta_b$ is imaginary. However, we know that $\beta_b$
is also real. Therefore, the only possible solution is that $\beta_b = 0$.   
Substituting this back in (\ref{co}) yields $\alpha_a = \alpha_b$.
We have thus, rederived the restrictions
$ \alpha_a = \alpha_b $ and $\beta_b = 0$ from the principle of 
non-increase of entanglement under LOCC. 

In conclusion, we have shown, by considering two different 
physical principles, that the Hadamard 
transformation for completely arbitrary unknown qubits is an invalid 
operation.
The Hadamard gate exists if and only if the state belongs to the 
special ensemble (\ref{rq}).   
Interestingly, quantum mechanics and no-signalling condition gives 
exactly the same set of states for a valid Hadamard operation. This bears 
similarity with optimal fidelity of a qubit \cite{bh, gisin}, the value 
of which turns out to be exactly the same by these two different 
considerations. Since
the Hadamard gate, just 
like the NOT gate, is allowed for a certain class of states
on the bloch sphere, it 
imposes a weaker constraint on the quantum system than the no-cloning and 
no-deleting operations. 

It is a pleasure to thank G. Kar for useful discussions.


\begin{thebibliography}{10}

\bibitem{d}
D. Deutsch,
\newblock {\em Proc. Royal Society of London A} {\bf 425}, 73, (1989).

\bibitem{dj}
D. Deutsch and R. Jozsa,
\newblock {\em Proc. Royal Society of London A} {\bf 439}, 553, (1992).  

\bibitem{sh}
P.W. Shor, 
\newblock{\em Proc. 35th Annual Symp. on Found. of Comp. Sc., IEEE Press,
Los Alamitos, CA} (1994).
 
\bibitem{pati1}
A.K. Pati,
\newblock {\em Phys. Rev. A} {\bf 66}, 062319, (2002).

\bibitem{mp}
A. Maitra and P. Parashar,
\newblock to appear in {\em IJQI} (2006), arXiv:quant-ph/0505068.

\bibitem{wz}
W.K. Wootters and W.H. Zurek, 
\newblock {\em Nature} {\bf 299}, 802 (1982).

\bibitem{pb1} 
A.K. Pati and S. Braunstein,
\newblock {\em Nature} {\bf 404}, 164 (2000).

\bibitem{gisin}
N. Gisin,
\newblock {\em Phys. Lett. A} {\bf 242}, 1 (1998).

\bibitem{hs}
L. Hardy and D.D. Song, 
\newblock {\em Phys. Lett. A} {\bf 259}, 331 (1999).

\bibitem{pati2}
A.K. Pati,
\newblock {\em Phys. Lett. A} {\bf 270}, 103 (2000).

\bibitem{pb2}
A.K. Pati and S. Braunstein,
\newblock {\em Phys. Lett. A} {\bf 315}, 208 (2003).

\bibitem{horo}
M. Horodecki, R. Horodecki, A. Sen(De)and U. Sen,
\newblock quant-ph/0306044.

\bibitem{bh}
V. Buzek, M. Hillery and R.F. Werner,
\newblock {\em Phys. Rev. A} {\bf 60}, R2626 (1999).

\bibitem{kar}
I. Chattopadhyay, S.K. Choudhary, G. Kar, S. Kunkri and D. Sarkar,
\newblock {\em Phys. Lett. A}{\bf 351}, 384 (2006).

\end{thebibliography}
\end{document}